\documentclass[12pt]{spieman}
\usepackage[]{graphicx}
\usepackage{setspace}
\usepackage{amssymb}
\usepackage{tocloft}
\usepackage{amsmath}
\usepackage{sistyle}
\usepackage{paralist}
\usepackage{wasysym}
\usepackage{upgreek}
\usepackage{mathrsfs}
\usepackage{gensymb}
\usepackage{multirow}
\usepackage{units}

\title{Line Spread Functions of Blazed Off-Plane Gratings Operated in the Littrow Mounting} 

\author{Casey T. DeRoo,\supscr{a} Randall L. McEntaffer,\supscr{a} Drew M. Miles,\supscr{a} Thomas J. Peterson,\supscr{a} Hannah Marlowe,\supscr{a} James H. Tutt,\supscr{a} Benjamin D. Donovan,\supscr{a} Benedikt Menz,\supscr{b} Vadim Burwitz,\supscr{b} Gisela Hartner,\supscr{b} Ryan Allured,\supscr{c} Randall K. Smith,\supscr{c} Ramses G{\"u}nther,\supscr{d} Alex Yanson,\supscr{d} Giuseppe Vacanti,\supscr{d} Marcelo Ackermann\supscr{e} }

\affiliation{\supscrsm{a}University of Iowa, Department of Physics \& Astronomy, 210 Van Allen Hall, Iowa City, Iowa, 52242, USA \\
\supscrsm{b}MPI f{\"u}r extraterrestrische Physik, Giessenbachstrasse 1, D-85748, Garching, Germany \\
\supscrsm{c}Harvard-Smithsonian Center for Astrophysics, 60 Garden Street, Cambridge, MA USA \\
\supscrsm{d}cosine Science \& Computing BV, J.H. Oortweg 19, 2333 CH Leiden, The Netherlands \\
\supscrsm{e}cosine Research BV, J.H. Oortweg 19, 2333 CH Leiden, The Netherlands
}

\cftpagenumbersoff{figure}
\cftpagenumbersoff{table} 
\begin{document} 
\maketitle 

\begin{abstract}

Future soft X-ray (10 -- 50 $\angstrom$) spectroscopy missions require higher effective areas and resolutions to perform critical science that cannot be done by instruments on current missions. An X-ray grating spectrometer employing off-plane reflection gratings would be capable of meeting these performance criteria. Off-plane gratings with blazed groove facets operated in the Littrow mounting can be used to achieve excellent throughput into orders achieving high resolutions. We have fabricated two off-plane gratings with blazed groove profiles via a technique which uses commonly available microfabrication processes, is easily scaled for mass production, and yields gratings customized for a given mission architecture. Both fabricated gratings were tested in the Littrow mounting at the Max-Planck-Institute for extraterrestrial Physics PANTER X-ray test facility to assess their performance. The line spread functions of diffracted orders were measured, and a maximum resolution of 800 $\pm$ 20 is reported. In addition, we also observe evidence of a `blaze' effect from measurements of relative efficiencies of the diffracted orders. 
\end{abstract}


{\noindent \footnotesize{\bf Address all correspondence to}: Casey T. DeRoo, University of Iowa, Department of Physics \& Astronomy, 210 Van Allen Hall, Iowa City, Iowa, 52242, USA; Tel: +1 319-355-1835; Fax: +1 319-335-1753; E-mail:  \linkable{casey-deroo@uiowa.edu} }
\section{Introduction}
\label{section:introduction}

Soft X-ray wavelengths (10 -- 50 $\angstrom$) are host to a number of transition lines helpful in characterizing astrophysical plasmas in energetic environments. Grating spectrometers are the instrument of choice for observing spectra in this bandpass and typically consist of three major components: a set of focusing optics, a grating array, and a detector array. The focusing optic collects light from the source and directs it towards a focus several meters down the optical axis. Instead of being allowed to reach the focus, however, the converging light is intercepted by an array of grating elements. The periodic structure present on the gratings diffracts the converging light based on wavelength. The diffraction pattern is then imaged with an array of detectors at the focal plane, and the source spectrum is reconstructed based on the observed diffraction pattern. 

Grating spectrometers are employed on currently operating missions like the \emph{Chandra X-ray Observatory} and \emph{XMM-Newton}. However, the science requirements of future X-ray spectrometers necessitate significant improvements in instrument performance. \emph{Arcus}, for example, is a proposed X-ray grating spectrometer to be mounted on the International Space Station requiring resolution \mbox{R $(\lambda/\Delta \lambda) >$ 2000} and effective area $>$ 400 cm$^2$ over the 21.6 -- 25 $\angstrom$ bandpass in order to perform its critical science \cite{Arcus_reference}. This represents a substantial improvement in both metrics over currently existing capabilities, and will require significant investment in enabling technologies. 

\begin{figure}
\begin{center}
\begin{tabular}{c}
\includegraphics[height=8.5cm]{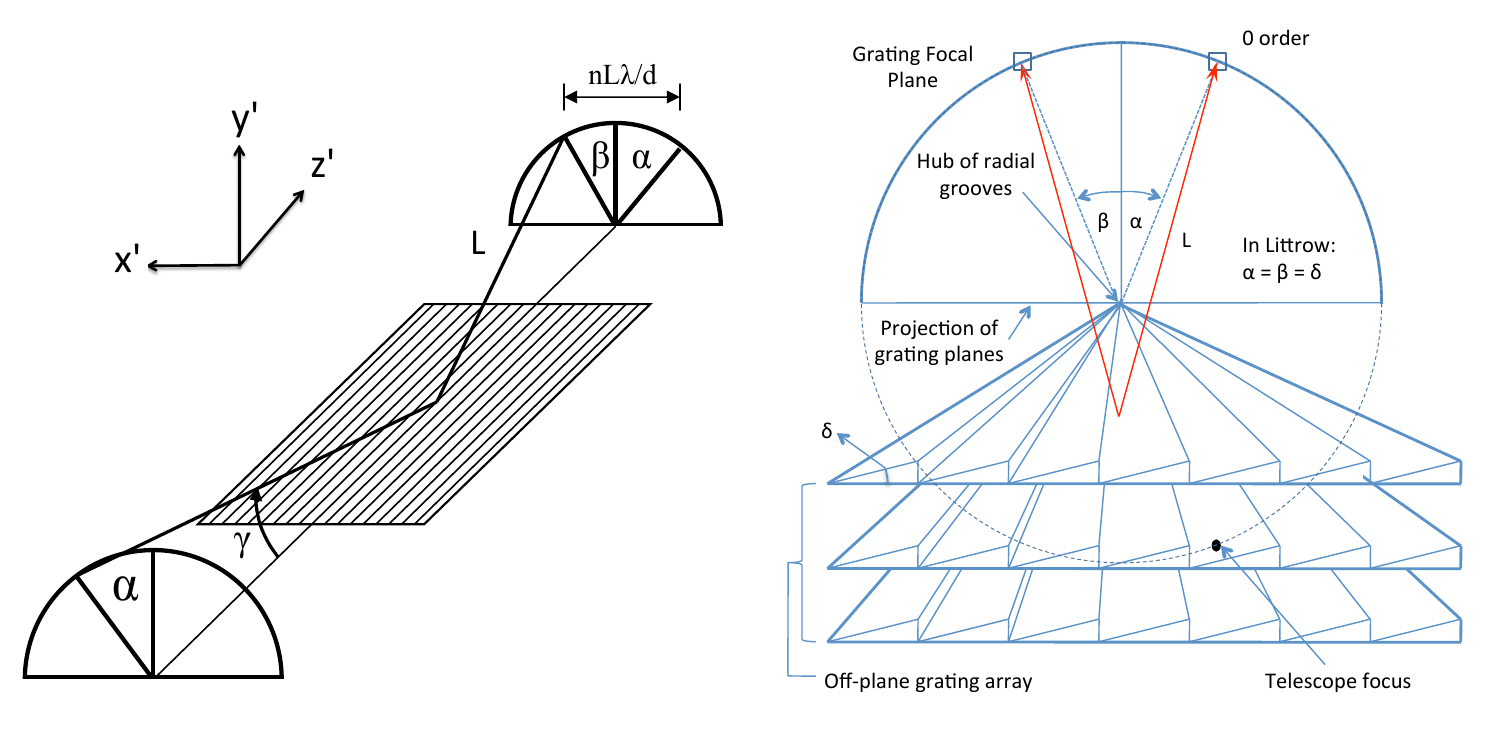} 
\\
(a) \hspace{8.5cm} (b)
\end{tabular}
\end{center}
\caption 
{\label{fig:diffraction_geometry} 
The diffraction geometry of off-plane gratings. (a) Light incident on the grating is parameterized via the half-cone angle $\gamma$ and a rotation angle $\alpha$ about the groove direction \emph{z'}. Diffracted light is constrained to the surface of a cone of the same half-cone angle $\gamma$, forming the arc of diffraction. (b) As seen in projection, light that would converge to the telescope focus is instead incident on an array of off-plane gratings. This light is then reflected to the $\alpha$ or diffracted to an angle $\beta$. This grating is operated in the Littrow mounting, where the $\alpha = \beta = \delta$, the facet angle.} 
\end{figure}

Off-plane reflection gratings are one such enabling technology, offering the ability to work at high dispersion while maintaining excellent throughput. In the off-plane mount (Figure \ref{fig:diffraction_geometry}), the grating grooves are oriented quasi-parallel to the direction of the incoming light. This geometry yields a diffraction pattern in which the outgoing orders are constrained to the surface of a half-cone, and hence is often referred to as `conical diffraction' . The grating equation for the off-plane mount is:
\begin{equation}
\sin{\alpha} + \sin{\beta} = \frac{n \lambda}{d \sin{\gamma}}\text{,}
\label{eq:grating_equation}
\end{equation}

\noindent{}where $d$ is the groove period, $\lambda$ is the wavelength of the diffracted light, $n$ is the order number, $\gamma$ is the half-cone opening angle between the incident beam and the groove direction, $\alpha$ is the azimuthal angle between the reflected (0\textsuperscript{th} order) spot and the grating normal as projected into the grating focal plane, and $\beta$ is the azimuthal angle between the diffracted spot and the grating normal projected into the grating focal plane. In this paper, we use a prime (') to denote the coordinate system defined by the grating, and define the grating focal plane to be the plane which is perpendicular to the direction of the central grating groove (\emph{z'}) and contains both the telescope focus and the 0\textsuperscript{th} order reflection.

While constrained to lie on the half-cone, the spectral information is contained in only one dimension, as can be shown by differentiating Eq. \ref{eq:grating_equation} with respect to the dispersion direction $x'$ $(= L \sin{\gamma} \sin{\beta})$:
\begin{equation}
\frac{\mathrm{d}\lambda}{\mathrm{d}x'} = \frac{10^7}{nL\mathrm{D}} \frac{\angstrom}{\mathrm{mm}}\text{.}
\label{eq:dispersion_equation}
\end{equation}
\noindent{}Here, $D$ is the groove density ($\equiv 1/d$) and $L$ is often referred to as the throw, and sets the size scale of the system. Thus, by Eq. \ref{eq:dispersion_equation}, the spectral width of a line is measured by its physical extent in the dispersion direction, and the resolution of a spectrometer governed by the width of the diffracted spot.

Akin to in-plane reflection gratings, off-plane gratings can be `blazed' to achieve maximum diffraction efficiency at a given wavelength. This geometric effect is brought about under a specific mounting condition, known as the Generalized Mar\'{e}chal and Stroke (GMS) mounting \cite{Neviere_1978_Rapid_Communication} or the off-plane Littrow mounting. The off-plane Littrow mounting is realized when $\alpha = \beta = \delta$, where $\delta$ is the facet angle of the grating. In this mounting, off-plane gratings are theoretically capable of achieving diffraction efficiencies approaching the reflectivity of the grating material \cite{Neviere_1978}.

In this paper, we present first results from two blazed off-plane reflection gratings made via a novel fabrication method capable of producing high-performance flight gratings. The fabricated gratings were tested at the Max Planck Institute for extraterrestial Physics (MPE) PANTER X-ray test facility. A silicon pore optics (SPO) stack was used in conjunction with the gratings to form a spectroscopic system, and the line spread functions (LSFs) of the diffracted orders were measured in order to assess grating performance. The fabrication requirements of high-performance off-plane gratings are explained in Sec. \ref{subsec:fab_requirements} and the manufacturing method used for the gratings tested here is described in Sec. \ref{subsec:fab_procedure}. An overview of the experimental set-up is given in Sec. \ref{section:PANTER}, the details of placing the grating into the Littrow mounting in Sec. \ref{subsec:measurement_configurations}, and a walkthrough of the data reduction process in Sec. \ref{subsec:data_reduction}. A discussion of the results of the test campaign are presented in Sec. \ref{section:results}. The significance of the work performed here, as well as a brief outline of work to be performed in the future, are discussed in Sec. \ref{section:conclusions}. 

\section{Off-Plane Grating Fabrication}
\label{section:fabrication}

\subsection{Fabrication Requirements}
\label{subsec:fab_requirements}
Meeting the performance specifications of future spectrometers like \emph{Arcus} requires a high performance diffraction grating -- that is, a grating capable of 40 -- 60\% throughput while operating at R $>$ 2000. To achieve optimal resolution and throughput, off-plane gratings require a customized facet and ruling geometry. First, the grating grooves must be radially ruled in order to realize high resolution \cite{Cash_1983}. This radial `fanning' of the grooves matches the convergence of the incident beam and ensures the inherent point spread function (PSF) of the telescope is not aberrated. The facets of the grating grooves must also be specially shaped to realize the off-plane blaze condition. A grating with a triangular groove profile placed in the Littrow mounting realizes high diffraction efficiencies for a segment of the diffraction arc near the direction of the facet normal (See Figure \ref{fig:diffraction_geometry}b). This segment is equivalent to a range of wavelengths for a given order. Hence, by tuning the facet angle during manufacture, the blaze effect can be used to increase a spectrometer's effective area near particular lines of interest.

In order to diffract at X-ray wavelengths, the groove densities for off-plane gratings must be large compared to in-plane diffraction gratings. Typical groove densities for off-plane gratings range from 4,000 -- 10,000 grooves/mm. The groove pattern must be producible over large formats ($\sim$100 cm\textsuperscript{2}) in order to achieve adequate geometric throughput at grazing incidence. Finally, the manufactured gratings need to meet the figure requirements for the spectrometer in question. Any grating substrate deviations from flat translate into local variations of grating orientation. These local variations blur the LSF at the focal plane and compromise overall instrument performance. Grating figure tolerances can be derived by considering the effect of grating misalignments on instrument performance \cite{Allured_2013}. In sum, the gratings for a high performance spectrometer should have: 
\begin{inparaenum}[\itshape a\upshape)]
\item radially fanned grating grooves,
\item blazed facets,
\item high groove densities,
\item large patterned areas, and
\item optical figure quality.
\end{inparaenum}

\subsection{Fabrication Procedure}
\label{subsec:fab_procedure}
\begin{figure}[!b]
	\begin{center}
	\begin{tabular}{c}
	\includegraphics[height= 6in]{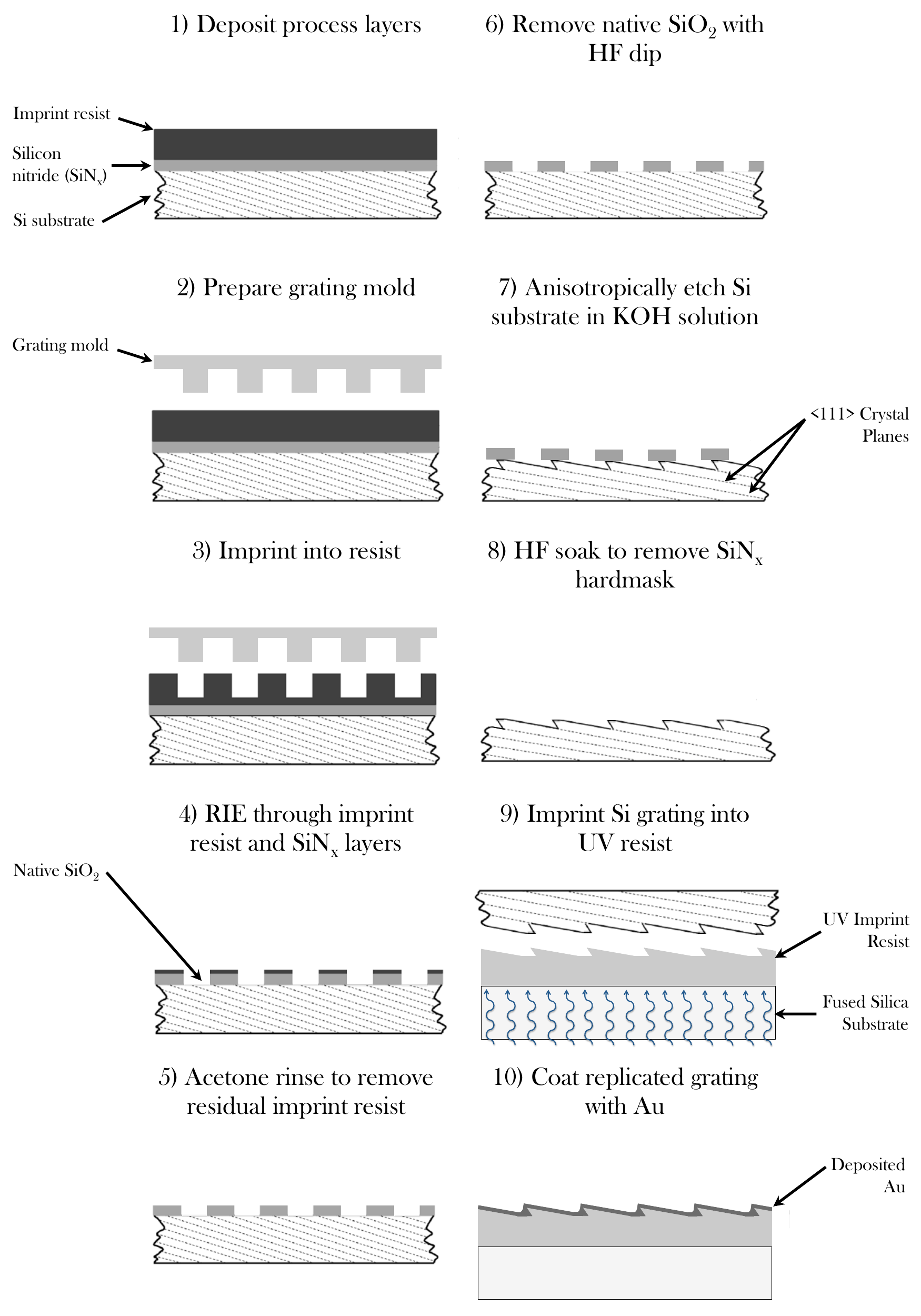}
	\end{tabular}
	\end{center}
	\vspace{-1.0em}
	\caption{\label{fig:fab_diagram} Procedure for fabricating off-plane gratings.}  
\end{figure}

Off-plane gratings meeting all of these requirements can be manufactured via a microfabrication process outlined in Figure \ref{fig:fab_diagram}. The fabrication procedure outlined here builds on work first performed by Franke et al. 1997\textsuperscript{ \cite{Franke_1997}}, who manufactured an X-ray reflection grating using an anisotropic potassium hydroxide (KOH) etch to form blazed facets in a silicon substrate. A further iteration of blazed X-ray reflection grating fabrication was performed by Chang et al. 2003\textsuperscript{ \cite{Chang_2003}} and Chang et al. 2004\textsuperscript{ \cite{Chang_2004}}, who employed UV nanoimprint lithography (UV-NIL) to replicate a blazed silicon master also produced using the KOH etch technique. However, both of these authors employ interference lithography to create gratings that are straight ruled, rather than the radial rule required for high performance in the off-plane mounting. The grating fabrication process outlined in the present work uses electron beam lithography (EBL) and deep UV (DUV) projection lithography to produce a radially ruled grating `pre-master' \cite{McEntaffer_2013}. The resulting pre-master then serves as the mold in a thermal nanoimprint lithography (T-NIL) patterning step, creating a radially ruled etch mask for a subsequent KOH etch step. Thus, the technique presented here builds on these previous works, combining the flexibility of patterning with EBL and DUV projection lithography, the anisotropy of KOH etch technique, and the ease of replication offered by UV-NIL to produce radially ruled, blazed X-ray reflection gratings that can be made in large numbers. 

We also demonstrate the use of silicon wafers with crystallographic orientations besides \textless111\textgreater{} and \textless100\textgreater{} in order to manufacture gratings with facet angles near those proposed for off-plane X-ray spectrographs. In grating fabrication processes with a KOH etch step, the blaze angle of the grating is set by the angle between the [111] direction and the wafer normal. The grating grooves are then patterned parallel to the [01$\bar{\text{1}}$] direction in order to bound the KOH etch by \{111\} planes. Franke et al. 1997\textsuperscript{ \cite{Franke_1997}} and Chang et al. 2004\textsuperscript{ \cite{Chang_2004}} set the facet angle of their fabricated gratings by dicing a \textless111\textgreater{} ingot off-axis, while Chang et al. 2003\textsuperscript{ \cite{Chang_2003}} used both \textless100\textgreater{} and off-axis cut \textless111\textgreater{} Si wafers. However, silicon foundaries are typically only capable of realizing off-axis cuts of $<$ 10$\degree$, well below the blaze angles proposed for notional off-plane X-ray spectrographs. By way of an example, \emph{Arcus} requires a blaze angle near 30$\degree$, while the Off-Plane Grating Rocket Experiment (OGRE) proposes to use gratings with a 28$\degree$ blaze \cite{DeRoo_SPIE_2013}. By using other commercially available crystallographic orientations of silicon, however, a variety of blaze angles can be obtained. The resulting facet angles for a number of common silicon wafer crystallographic orientations are given in Table \ref{tab:crystallographic_blazes}. Furthermore, through off-axis cutting of wafers with different crystallographic orientations, it is possible to obtain any blaze angle that might be desired for an off-plane grating spectrometer in the future. 

\begin{table}[!b]
\begin{center}
	\caption{\label{tab:crystallographic_blazes}Facet angles achievable using different crystallographic orientations of silicon wafers.}
	\vspace{1.0em}
	\begin{tabular}{| c | c |}
		\hline
		Wafer Orientation & Blaze Angle \\ \hline \hline
		\textless111\textgreater & $0\degree$\\ \hline
		\textless211\textgreater & $22.4\degree$ \\ \hline
		\textless311\textgreater & $29.5\degree$ \\ \hline
		\textless511\textgreater & $38.9\degree$ \\ \hline
		\textless711\textgreater & $43.3\degree$ \\ \hline
		\textless100\textgreater & $54.7\degree$ \\
		\hline
	\end{tabular}
\end{center}
\end{table}

In the interest of completeness, the specifics of the fabrication process employed are detailed here. First, a silicon wafer of the desired orientation is coated with two process layers: a 30 nm layer of silicon nitride (SiN$_x$) deposited via low-pressure chemical vapor deposition (LPCVD), and 95 nm thick layer of NXR-1025 nanoimprint resist deposited via spin coater (Figure \ref{fig:fab_diagram}, Step \#1). These layers are deposited over the thin native silicon dioxide layer present on the substrate. Next, a grating pre-master with the desired groove distribution is obtained and prepared for use (Step \#2). The pre-master is a grating which has the desired groove density and radial convergence, is identical in size to the final flight gratings, and will serve as a mold for the nanoimprint process. However, the pre-master has a laminar (i.e. `square wave') groove profile and lacks the figure quality required for flight gratings. Prior to use, the pre-master is coated with a mold release agent to aid in separation of the mold from the substrate after imprinting. The pre-master is then aligned to a foundry-provided wafer flat indicating the [01$\bar{\text{1}}$] direction and the grating pattern imprinted into the resist using a Nanonex NX-1006 nanoimprint tool (Step \#3). Any nanoimprint resist remaining in the groove troughs is then etched with a reactive ion etch (RIE) in Ar/O$_2$ plasma performed at 10 mTorr and 40 W RF, and the SiN$_x$ layer etched in a O$_2$/CHF$_3$ plasma at 100 mTorr and 150 W RF (Step \#4). A rinse step (Step \#5) in acetone removes any remaining nanoimprint resist, leaving a silicon nitride hardmask matching the grating mold pattern in negative. A dip in buffered HF (Step \#6) removes the native layer of silicon dioxide, exposing bare silicon between strips of the nitride hardmask. The sample is then transferred to a chemical bath for an anisotropic potassium hydroxide (KOH) wet etch (Step \#7) to sculpt the triangular shape of the groove facets. After terminating the KOH etch with a brief soak in DI water, the silicon nitride mask is removed by a soak in hydrofluoric acid (HF) (Step \#8). 

At this point, the only requirement listed in Section \ref{subsec:fab_requirements} not met by the existent sample is optical figure quality. Silicon wafers have global flatness specifications that are outside the figure qualities needed for off-plane gratings \cite{Allured_2013}. Fused silica substrates, on the other hand, can be manufactured to be optically flat to high precision at reasonable cost. By imprinting the blazed silicon grating into resist on a fused silica substrate (Step \#9), the radially ruled, blazed grating profile can be replicated on a surface meeting the required figure. UV nanoimprint lithography (UV-NIL) is employed for this final replication step. As a secondary benefit, a second imprint makes the production of flight gratings a more cost- and time-efficient process, as the same silicon grating can be used for multiple imprints, boosting process yield. The deposition of a thin, X-ray reflective layer over the fabricated grating (Step \#10) then yields an off-plane grating meeting all the fabrication requirements described in Sec. \ref{subsec:fab_requirements}. 

\section{PANTER Test Assembly}
\label{section:PANTER}
Two gratings with different facet angles were fabricated using the anisotropic KOH wet etch method described in Sec. \ref{subsec:fab_procedure} and tested at the PANTER X-ray test facility \cite{PANTER_reference}. The PANTER facility consists of several X-ray sources housed at one end of a \mbox{120 m} long, 1 m diameter vacuum chamber. This forms a long beamline, limiting the angles of divergence from the X-ray source and resulting in a quasi-collimated beam. At the opposite end, a 12 m long, 3.5 m diameter instrument chamber is joined to the beamline and houses several customizable optical benches which can be maneuvered with vacuum stages. 

\begin{figure}[b]
	\begin{center}
	\begin{tabular}{c}
	\includegraphics[width=6.0in,clip = true, trim = 0.0in 0.0in 0.0in 0.5in]{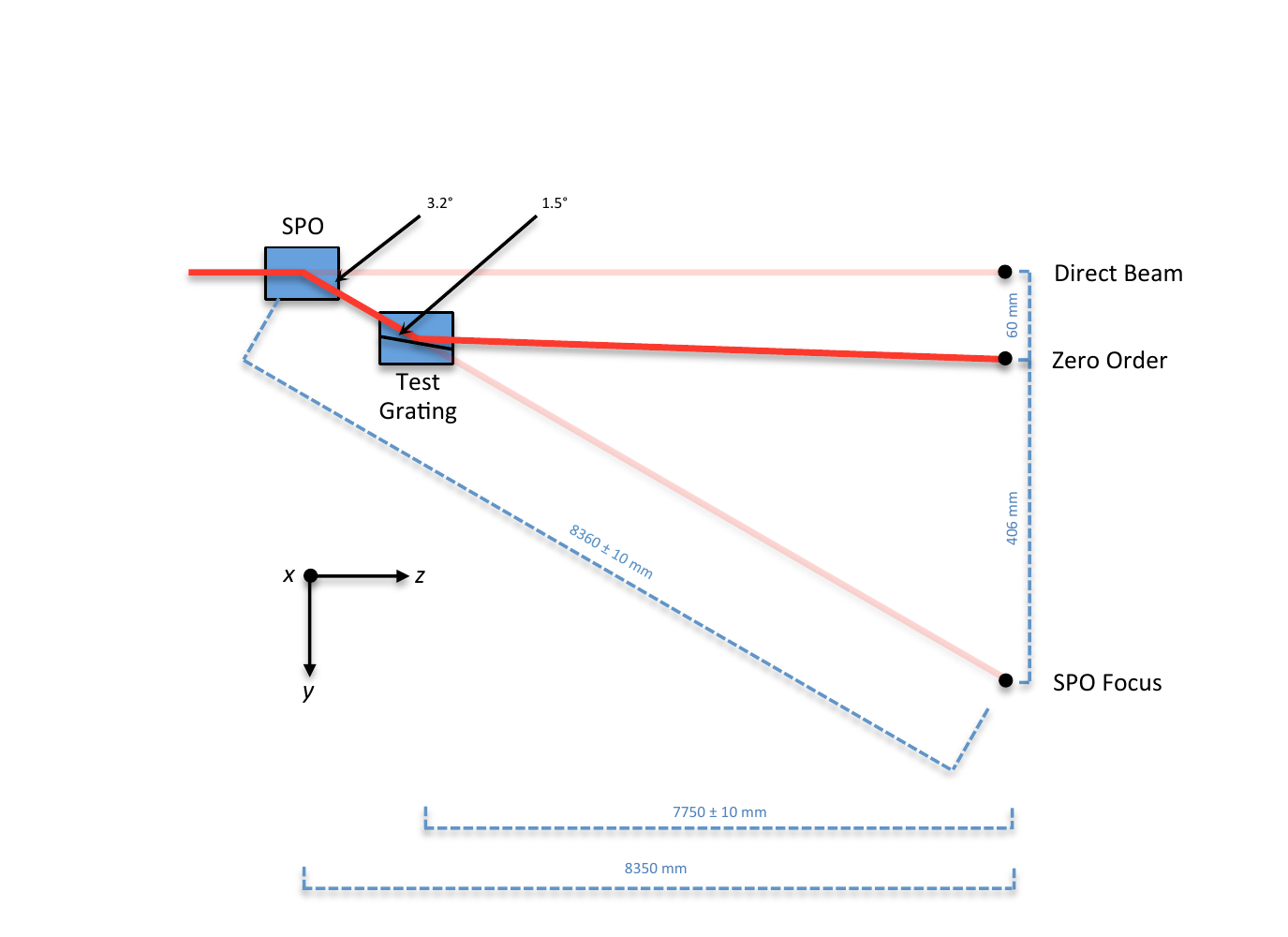}
	\end{tabular}
	\end{center}
	\caption{\label{fig:PANTER_Setup} A diagram showing the optical configuration of the off-plane grating test setup at PANTER. The coordinates specified here (\emph{x}, \emph{y}, and \emph{z}) are system coordinates and are aligned to the grating coordinate system given in Figure \ref{fig:diffraction_geometry} with mechanical tolerances. Distances with quoted errors were directly measured using a laser distance meter, while distances without errors are inferred from the SPO incidence angle and grating graze angle $\eta$. } 
\end{figure}

The off-plane grating test assembly employed at the PANTER facility for this set of tests consisted of a Silicon Pore Optics (SPO) stack which serves as a focusing optic, an off-plane grating mechanical interface affixed to an optical bench capable of changing the mounting of a grating \emph{in situ}, and a suite of X-ray detectors to sample the diffraction pattern at the focal plane. A diagram showing the relative positions of the components is shown in Figure \ref{fig:PANTER_Setup}. An electron impact source with a Mg target and a 12.5 $\upmu$m thick filter was used to generate the X-ray flux. The Mg K$\alpha$ line is composed of two primary components, Mg K$\alpha_1$ and Mg K$\alpha_2$, separated by 0.265 eV at a 2:1 intensity ratio \cite{Klauber_1993}. For the purposes of this paper, we refer to these lines together as the Mg K$\alpha$ line with a central wavelength of 9.8900 \angstrom{} \cite{Bearden_1967}. 

\subsection{Silicon Pore Optics}
\label{subsec:SPO}
SPO \cite{SPO_Ref1} have been developed for the past 10 years by a consortium led by cosine Research, and have become the main technology for the X-ray mirrors of the Athena mission \cite{SPO_Ref2}. SPO are formed from highly polished silicon wafers which are diced into a rectangular shape and ribbed, leaving a thin membrane on one side used to reflect the X-rays, and a number of ribs on the opposite side that are used to bond to the next plate. A series of plates are then bent and stacked to form pores which permit X-rays to reflect and pass through to the focal plane. By elastically deforming the plates around a conical mandrel and wedging each plate to give a small angular offset matching the change in radial position, the mirror plates can be bent into a conical approximation of paraboloids or hyperboloids, thus enabling the construction of a stiff, lightweight, high resolution imaging system.

For this campaign, a single SPO stack was built. The geometry of the stack approximates the geometry of a parabolic reflector. The stack consists of 13 plates, with radii between 439 mm and 450 mm, width of 66 mm, and axial length of 22 mm. The constructed SPO realizes its best focus at an axial distance of 8350 $\pm$ 10 mm and is constructed such that the incidence angle for on-axis measurements is 1.6$\degree$.

\begin{figure}
	\begin{center}
	\begin{tabular}{c}
	\includegraphics[height=5.5cm]{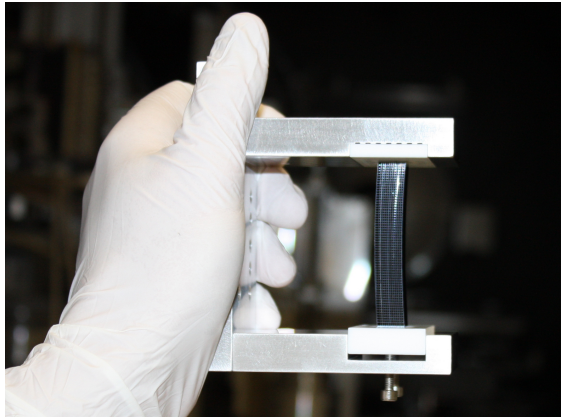}
	\end{tabular}
	\end{center}
	\caption{\label{fig:SPO_stack_image} A image of the SPO stack before being installed in the PANTER X-ray test chamber.} 
\end{figure}

In terms of mirror geometry, the constructed SPO stack is similar to the primary SPO stack for the proposed \emph{Arcus} mission. However, due to time and budget constraints, the methods employed to build this particular stack are atypical of SPO production. First, this stack was shaped on a simple aluminum mandrel rather than one of high quality polished silicon. In addition, the stacking device employed to deform the plates around the mandrel was not retooled to accomodate the change in radii from the nominal SPO stacking radius of 800 mm. Thus, the performance of the SPO module used in these tests is not representative of the state of the art in SPO manufacture. An image of the SPO stack prior to installation in the chamber is shown in Figure \ref{fig:SPO_stack_image}. 

A series of aperture masks mounted to vacuum stages were used to control the illumination of the SPO by the source. For the present test of blazed gratings, a single mask measuring 42 mm by 11 mm was positioned in front of the SPO, subaperturing the optic to this width and radial extent respectively. This mask represents an effective SPO illumination percentage of 63\%, and will hence be referred to as the 63\% mask. However, during initial characterization of the optic, a mask measuring 6 mm by 11 mm (representing an effective illumination percentage of 9\%, and hence dubbed the 9\% mask) was also employed to initially characterize the SPO. 
 
\subsection{Detectors}
\label{subsec:detectors}
Three focal plane instruments were employed to sample the SPO focus and the arc of diffraction. The Position Sensitive Proportional Counter (PSPC) is a gas-proportional counter and is the flight spare of detector onboard ROSAT \cite{PSPC_reference}. Though the PSPC has relatively low spatial resolution, the large format of the detector ($\diameter =$ 80 mm) is useful for viewing large portions of the arc of diffraction at once and establishing rough alignment. A CCD camera is then used to image diffracted orders with higher spatial resolution. The Third Roentgen Photon Imaging Camera (TRoPIC) is a back-illuminated CCD and is a smaller version of the detector baselined for the eROSITA mission \cite{TROPIC_reference}. The second CCD camera in use is the Princeton Instruments X-ray Imager (PIXI), a soft X-ray imager with a pixel size of 20 $\upmu$m. PSPC and TRoPIC were mounted just above the nominal horizontal (\emph{y} in Figure \ref{fig:PANTER_Setup}) system axis, while PIXI was mounted just below. The suite of detectors is capable of $\pm$250 mm of motion along the optical axis (\emph{z} in Figure \ref{fig:PANTER_Setup}) and has enough travel to cover the entire grating focal plane in the \emph{y} direction and a total of 150 mm in the \emph{x} direction. At the nominal grating graze angle of 1.5$\degree$, the arc of diffraction is too large to be fully sampled in the dispersion direction, and thus some diffracted orders were not accessible to any detector. The nominal position of the detectors was chosen to allow TRoPIC to sample the SPO focus, the reflected beam (0\textsuperscript{th} order), some positive orders and limited negative orders and allow PIXI to sample the SPO focus. (See Figures \ref{fig:Order_Location_Grating1}, \ref{fig:Order_Location_Grating2}). 

\subsection{Gratings}
Both fabricated gratings tested at PANTER were patterned using a pre-master identical to the gratings tested in \cite{McEntaffer_2013}, which have a nominal groove density of 6033 gr/mm, a radial convergence matching a 8.4~m focal length telescope, and a format measuring 25 mm $\times$ 32 mm, where the long edge is parallel to the ruling direction. The first grating (for ease of reference, \mbox{Grating 1}) was made using a \textless111\textgreater{} silicon wafer which was cut 10$\degree$ off-axis, forming a 10$\degree$ angled facet. This profile was successfully transferred to a 4~in. fused silica wafer and coated with a 10 nm thick layer of gold. The second grating was made using a \textless311\textgreater{} silicon wafer (Grating 2), yielding a 29.5$\degree$ facet angle, similar to the facet angles required for several future spectrometers including \emph{Arcus} and the Off-Plane Grating Rocket Experiment (OGRE) \cite{DeRoo_SPIE_2013}. Due to time and budget constraints, the blazed silicon grating was tested directly rather than transferring the grating pattern to a fused silica wafer, and was cleaved to measure approximately 35 mm by 45 mm (where the long edge is again parallel to the ruling direction) for mounting purposes.  

These gratings were mounted on a vacuum stage stack with four degrees of freedom: linear motion in the cross-dispersion direction \emph{y}, linear motion in the dispersion direction \emph{x}, grating graze angle $\eta$ (rotation about \emph{x}), and grating yaw $\Psi$ (rotation about \emph{y}). The grating testing positions were found prior to evacuating the chamber via the use of an optical laser mounted at the source end of the beamline. The laser spot could be passed through the SPO, and each grating maneuvered via vacuum stages until illuminated by the focused beam. The nominal grating graze angle of 1.5$\degree$ was set by measuring the distance between the reflected spot and the direct SPO focus as created by this same optical laser, and has an accuracy of 4$'$ given the uncertainty in the measured distance between the SPO stack and the detector plane ($\pm$10 mm over 8360 mm). Finding the zero yaw position of each grating was performed under X-ray illumination. Deviations from zero yaw move the groove `hub,' or point at which the radial grooves converge, increasing the radius of the diffraction arc without changing the zero order position. A yaw misalignment thus causes positive and negative orders to appear at different cross-dispersion coordinates relative to 0\textsuperscript{th} order. To perform the initial yaw alignment, PSPC was employed to examine large sections of the diffraction arc until positive and negative orders were symmetrically distributed about 0\textsuperscript{th} order. TRoPIC measurements were then used to perform a fine yaw alignment by centroiding $\pm$1\textsuperscript{st} Mg K$\alpha$ orders and adjusting the grating yaw until both orders were measured to be at the same cross-dispersion pixel value. Changes in centroid height were easily distinguishable with yaw rotation step sizes of 1.5$'$. We therefore estimate the accuracy of this method of yaw alignment to be 0.75$'$.

\section{Measurements}
\subsection{Test Configurations}
\label{subsec:measurement_configurations}
Following the characterization of each grating's orientation with respect to the stage axes, each grating was then placed in the Littrow mounting. Functionally, reaching the Littrow mount involves setting the graze angle $\eta$ and grating yaw $\Psi$ around the fixed geometry of the beamline and SPO: increasing the graze angle serves to increase the length of the chord between the SPO focus and zero order, while increasing the yaw of the grating increases both the radius of the diffraction arc and $\alpha$ in the grating equation. Recall that in the Littrow configuration, $\alpha = \beta = \delta$. Thus, the blaze wavelength $\lambda_b$ for a given order $n$ is the wavelength diffracted to an angle $\beta = \delta$ and is given by the expression:
\begin{equation}
\lambda_{b} = \frac{2 d\, \sin{\gamma}\sin{\delta}}{n}\text{,}
\label{eq:blaze_wavelength}
\end{equation}
\noindent{}which can be derived from Eq. \ref{eq:grating_equation}. The graze angle of the grating can thus be chosen to place a \mbox{Mg K$\alpha$} order at the blaze wavelength. Once the graze angle is set, the grating is then placed into the Littrow mounting by `yawing' the grating until the following relationship is satisfied: 
\begin{equation}
\sin{\Psi} = \tan{\eta}\tan{\delta}\text{.}
\label{eq:littrow_mounting_relationship}
\end{equation}
\noindent{}This relationship describing the Littrow mount is independent of the length scale of the system and can be derived from geometrical considerations. 

\begin{figure}[!btp]
	\begin{center}
	\begin{tabular}{c}
	\includegraphics[height=3.0in]{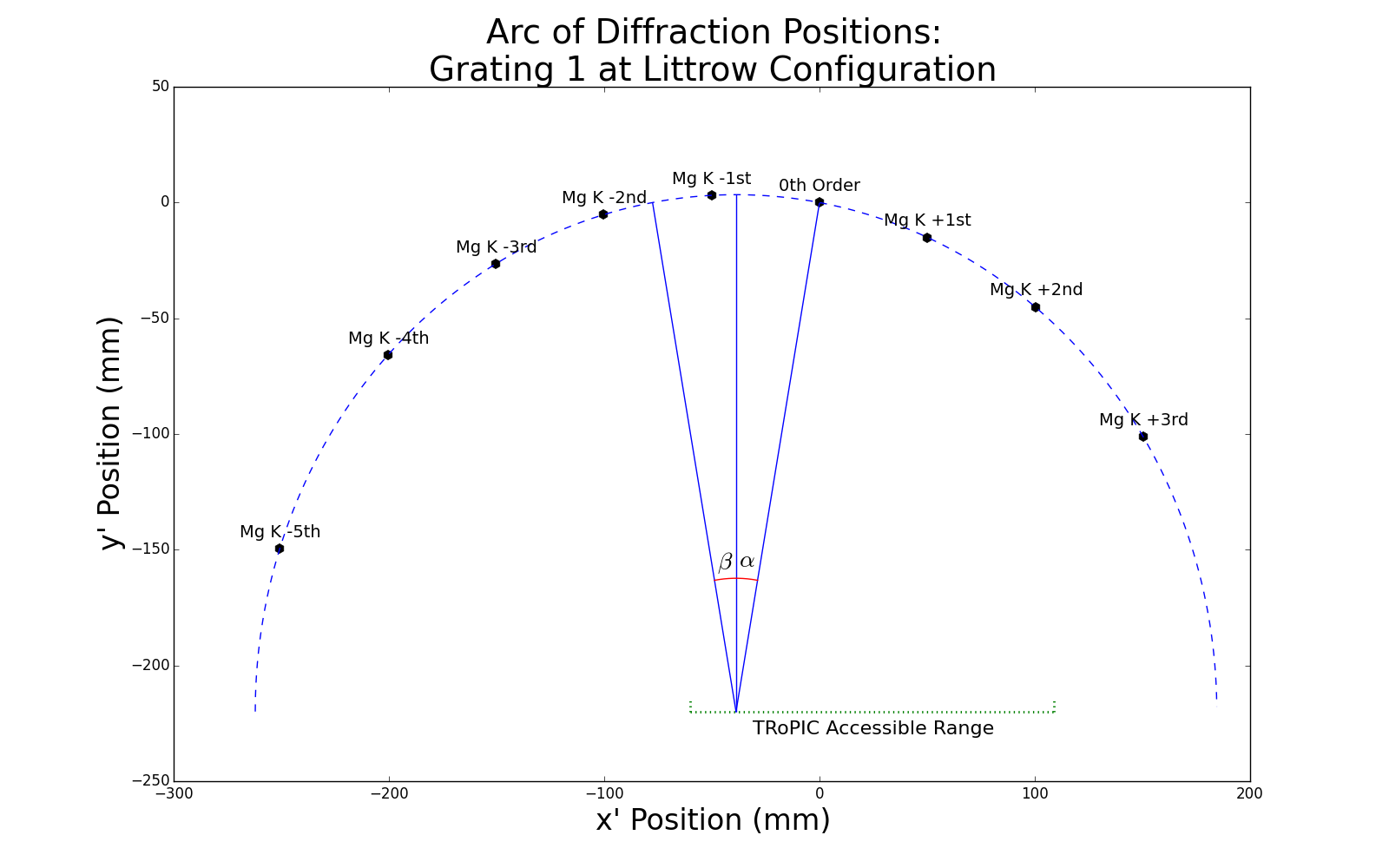}
	\end{tabular}
	\end{center}
	\caption{\label{fig:Order_Location_Grating1}A nomogram of the diffraction arc for Grating 1 in the Littrow mounting described. As the grating has a facet angle of $\delta = 10\degree$, placing the grating in the Littrow mount means that $\alpha = \beta = 10\degree$.} 

	\begin{center}
	\begin{tabular}{c}
	\includegraphics[height=3.0in]{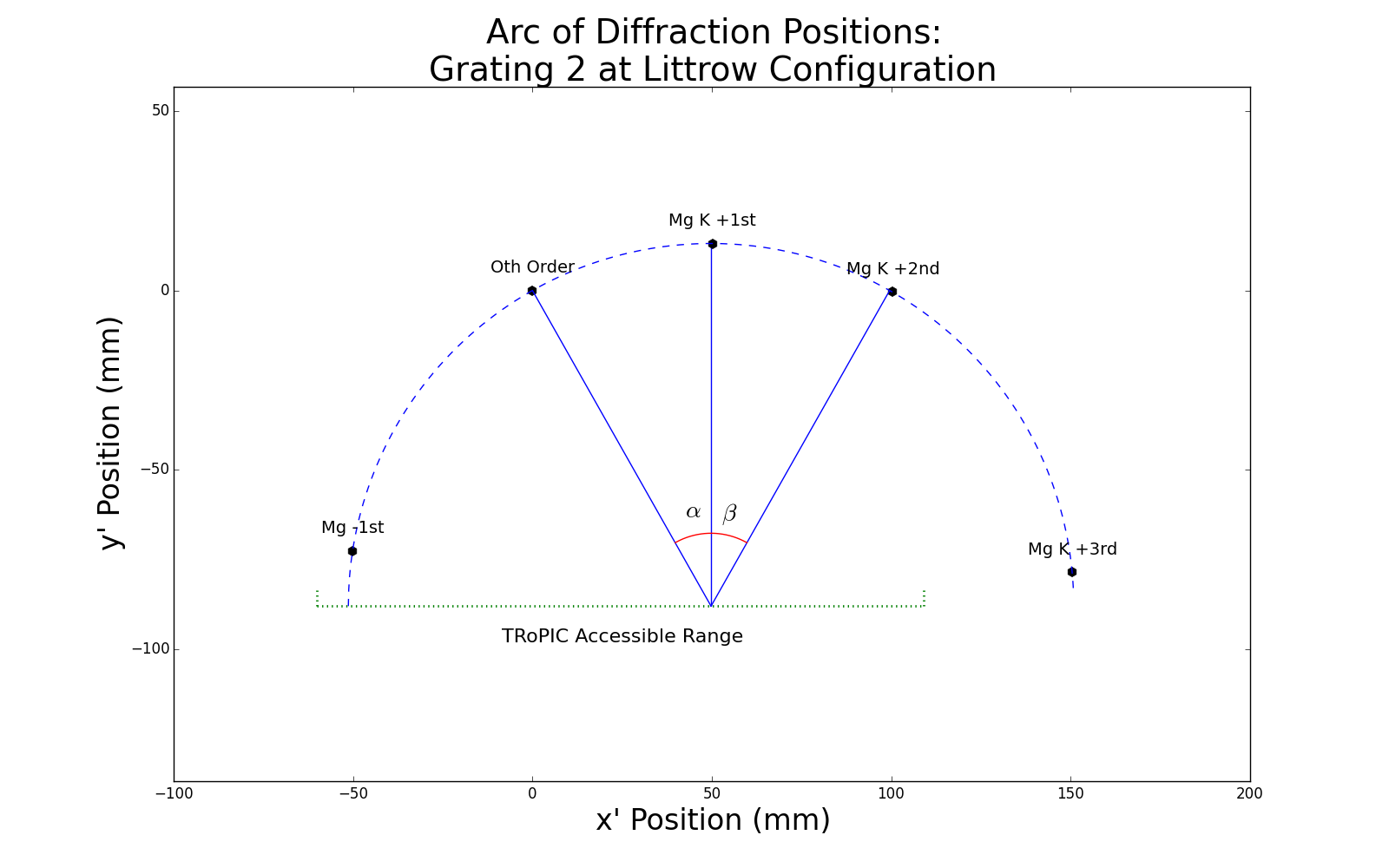}
	\end{tabular}
	\end{center}
	\caption{\label{fig:Order_Location_Grating2}A nomogram of the diffraction arc for Grating 2 in the Littrow mounting described. In the Littrow mounting, $\alpha = \beta = \delta$, and the facet angle $\delta = 29.5\degree$ for Grating 2.} 
\end{figure}

Figure \ref{fig:Order_Location_Grating1} shows the diffraction geometry of the test configuration for Grating 1. The groove facets of Grating 1 were blazed toward negative orders. A graze angle of $\eta$ = 1.5$\degree$ was chosen to be identical to the graze angle for \emph{Arcus} and OGRE. However, the blaze position does not correspond to the wavelength of any line fluoresced by the Mg target, lying approximately halfway between Mg K$\alpha$ -2\textsuperscript{nd} order and Mg K$\alpha$ -1\textsuperscript{st} order. For measurements of Grating 1, TRoPIC was used to image the Mg K$\alpha$ -1\textsuperscript{st}, +1\textsuperscript{st} order lines, as well as the 0\textsuperscript{th} order reflection.

The facet angle of $\delta$ = 29.5$\degree$ defines a different diffraction geometry for Grating 2 when placed in the Littrow mounting, which is shown in Figure \ref{fig:Order_Location_Grating2}. The groove facets of Grating 2 were blazed towards positive orders. Testing Grating 2 at a graze angle of $\eta = $ 1.5$\degree$ was precluded by the extent of detector travel: accessing the location of the ideal $\beta$ at such a graze angle required 250 mm of stage travel in the dispersion direction relative to 0\textsuperscript{th} order, which exceeds the maximum travel extent range of 150 mm as discussed in Sec. \ref{subsec:detectors}. Instead, a graze angle of 0.6$\degree$ was selected for testing this grating, which places the Mg K$\alpha$ +2\textsuperscript{nd} order line near the location of highest diffraction efficiency. For Grating 2, the Mg K$\alpha$ 0\textsuperscript{th} order, -1\textsuperscript{st}, +1\textsuperscript{st}, +2\textsuperscript{nd} diffraction orders were measured. 

The physical size of each grating relative to the aperture of the SPO means that each grating undersamples the PSF produced by the optic, even with the 63\% mask in place. As seen in projection, Grating 1 represents an effective aperture of 25 mm wide and 0.84 mm in radial extent (a 3\% effective illumination), while the Grating 2 subapertures the SPO to 25 mm by 0.33 mm (a 1\% effective illumination). As no attempt was made to align the gratings to an individual SPO plate, Grating 1 thus likely samples $\sim$2 SPO reflectors, while Grating 2 samples a single reflector. A summary of each grating's characteristics and mounting is provided in Table \ref{tab:test_outline}.

\begin{table}[!pb]
\begin{center}
	\caption{\label{tab:test_outline}An overview of the gratings tested at the PANTER X-ray facility, the test configuration for each grating, and the diffraction orders imaged.}
	\vspace{1.0em}
	\begin{tabular}{| c | c | c | c | c | c |}
		\hline
		\multirow{2}{*}{Grating} & Facet & Graze  & \# of SPO  & $\lambda_b$ & Mg K$\alpha$ Orders \\ 
		& Angle ($\delta$) & Angle ($\eta$) & Plates Sampled & ($n$ = 1) &  Imaged \\ \hline \hline
		Grating 1 & 10$\degree$ & 1.5$\degree$ & 1 -- 2 & 15.3 $\angstrom$ & -1\textsuperscript{st}, +1\textsuperscript{st}, 0\textsuperscript{th} \\ \hline
		Grating 2 & $29.5\degree$ & 0.6$\degree$ & 1 & 19.7 $\angstrom$ & -1\textsuperscript{st}, +1\textsuperscript{st}, +2\textsuperscript{nd}, 0\textsuperscript{th}\\ \hline
	\end{tabular}
\end{center}
\end{table}

\subsection{Data Reduction}
\label{subsec:data_reduction}
As implied by Eq. \ref{eq:dispersion_equation}, the spectral information of a diffracted order is contained only in \emph{x'}, the dispersion direction. Hence, the LSF is measured by collapsing the image in the cross-dispersion direction and measuring the total number of photons in each spatial bin. For measurements taken with TRoPIC, the CCD images are converted from native ADC units into individual photon events via event processing (`photon counting'), which is integrated into the onboard electronics. These output an event list characterized by detector coordinates, pulse height, and event time. A set of grading criteria is used to determine good events, and a split pixel analysis is then performed to give photon positions within subpixel accuracy. The resulting good event list is then minimally binned to the effective spatial resolution of 20 $\upmu$m \cite{TRoPIC_pixel_reference}. 

PIXI, on the other hand, has no integrated photon counting mode. Data taken with PIXI is reduced by the subtraction of a dark frame, where the dark frame is an array matching the format of the CCD whose values are a pixel-by-pixel average of a series of frames taken at an identical stage position with the X-ray source turned off. This same series is also used to construct an array called a variance frame, in which the array values are the pixel-by-pixel standard deviation $\sigma_{pix}$ of the series. The variance frame is then used to threshold the dark-corrected images by setting to zero any pixels falling below 3$\sigma_{pix}$. The resulting images are then in units of integrated ADC counts.

Next, a rotation is applied to the data to account for the misalignment of the dispersion direction to the horizontal detector axis. The dispersion direction \emph{x'} is set by the orientation of the grating and is perpendicular to both the groove direction and the grating normal. While the grating is nominally aligned to the detector axes via mechanical tolerances, the extent of the LSF in the \emph{y} direction means that a small angular misalignment between the detector axes and grating axes can yield a significant change in the measured width of the line. In order to account for this misalignment, the absolute position of each order in system coordinates (\emph{x}, \emph{y}) is first measured using the stage location and the LSF centroid as measured on the detector and defining the origin to be the 0\textsuperscript{th} order spot. The rotation angle required to convert system coordinates into grating coordinates (\emph{x'}, \emph{y'}) is then derived by enforcing the condition that the \emph{x'} positions of each diffracted order are appropriate integer multiples of one another, where the integer multiple is given by the known diffraction order. In this way, the order locations in the grating coordinate system are made to be self-consistent with periodic diffraction in \emph{x'}, e.g. the +2\textsuperscript{nd} order Mg K$\alpha$ line is twice as far in \emph{x'} from 0\textsuperscript{th} order as the +1\textsuperscript{st} order Mg K$\alpha$ line. The derived rotation angles are each found to be $<$ 1$\degree$, which is in keeping with the mechanical alignment tolerances of the detectors to the grating stage stack. These rotations are then applied to the data from each grating to yield CCD images where the horizontal detector axis is aligned to the dispersion direction.

\begin{figure}
	\begin{center}
	\begin{tabular}{c}
	\includegraphics[height=3.6in]{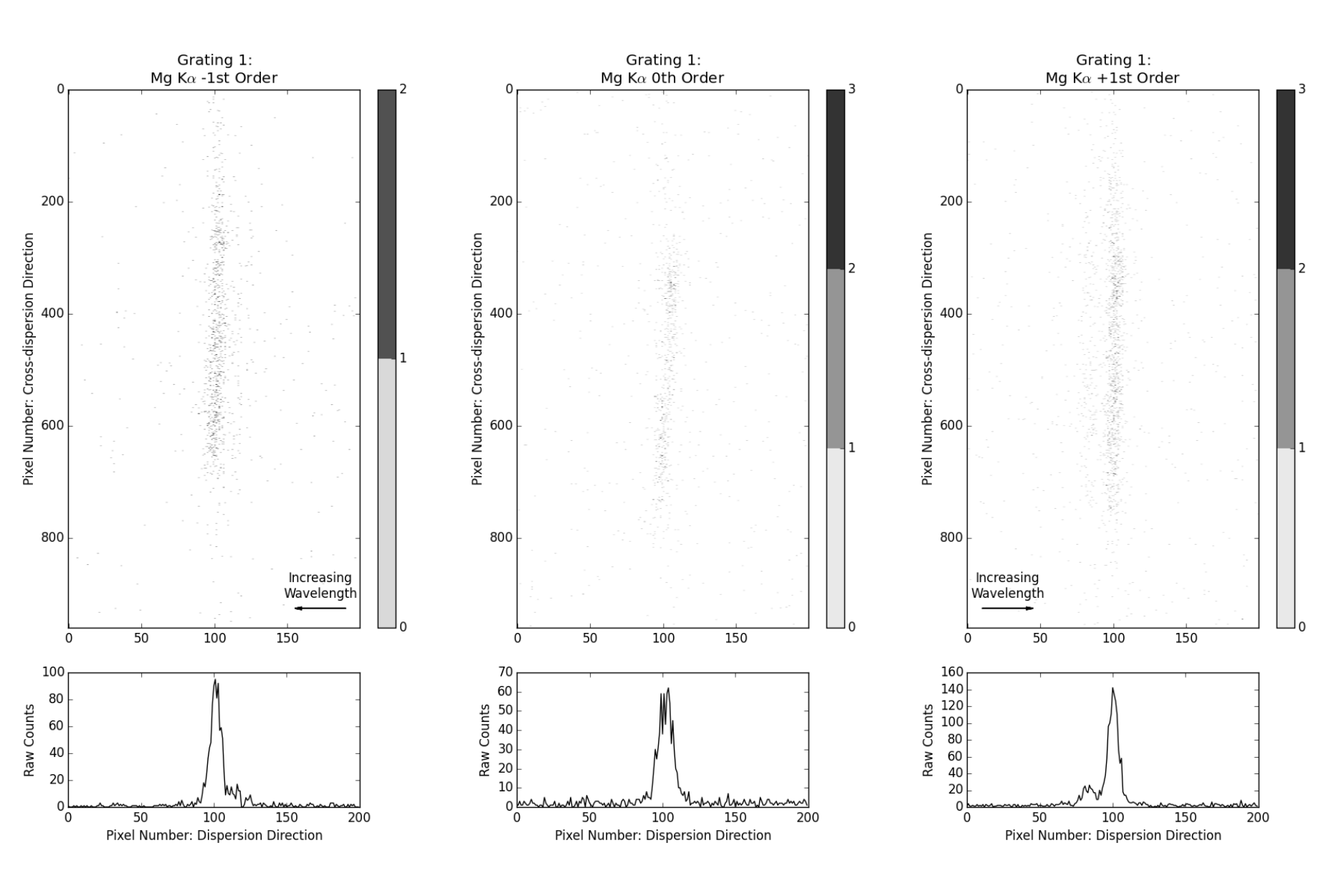}
	\\
	(a) \hspace{3.9cm} (b) \hspace{3.9cm} (c)
	\end{tabular}
	\end{center}
	\caption{\label{fig:example_CCD_image} TRoPIC images of (a) Grating 1 -1\textsuperscript{st} order, (b) Grating 1 0\textsuperscript{th} order, and (c) Grating 1 +1\textsuperscript{st} order lines. The insets below show the data collapsed in the cross-dispersion direction to yield the LSFs. A spectral feature consistent with the Mg K$\alpha_3$ satellite line (9.823 \angstrom{}) is visible to the right of the Mg K$\alpha$ line in (a) and to the left of the Mg K$\alpha$ line in (c). } 
\end{figure}

This reduction process yields a set of analysis images which are representative of the spectrometer configuration as tested. The images are subsequently summed in the cross-dispersion direction in order to measure the LSF. By way of an example, Figure \ref{fig:example_CCD_image} shows cropped analysis images and LSFs of Grating 1 Mg K$\alpha$ 0\textsuperscript{th} order, Mg K$\alpha$ +1\textsuperscript{st} order, and Mg K$\alpha$ -1\textsuperscript{st} order. Rather than modeling the observed LSFs with an expected functional form and extracting line widths from a model-dependent fit, we characterize each LSF by measuring its Half-Energy Width (HEW). The HEW provides an unambiguous, model-independent method of quantifying the spatial extent of flux concentrated in an X-ray feature. The HEW is found by constructing the Cumulative Distribution Function (CDF) of the LSF and calculating the spatial extent needed to enclose 50\% of the total number of photon events, $N_{tot}$. The upper and lower bounds on the HEW are calculated by determining the spatial extent needed to bound enough events to be within a single Poisson error of half of the total counts in a given line. As each sampled line has sufficient counts for the total number of counts over the given integration time to be normally distributed, we expect the Poisson counting error $\sqrt{N_{tot}}$ to converge to $\upsigma$, the standard deviation of the distribution of total counts over the given integration time. Hence, we refer to the calculated upper and lower bounds as 1$\upsigma$ bounds. It should be understood, however, that $\upsigma$ is in reference to the number of counts contained in the stated spatial extent and is not an error derived by fitting the observed LSF for the HEW.

\section{Results and Discussion}
\label{section:results}
\begin{figure}[!tbp]
	\begin{center}
	\begin{tabular}{c}
	\includegraphics[height=7.7in,clip=true,trim = 1.0in 1.4in 1.0in 1.0in]{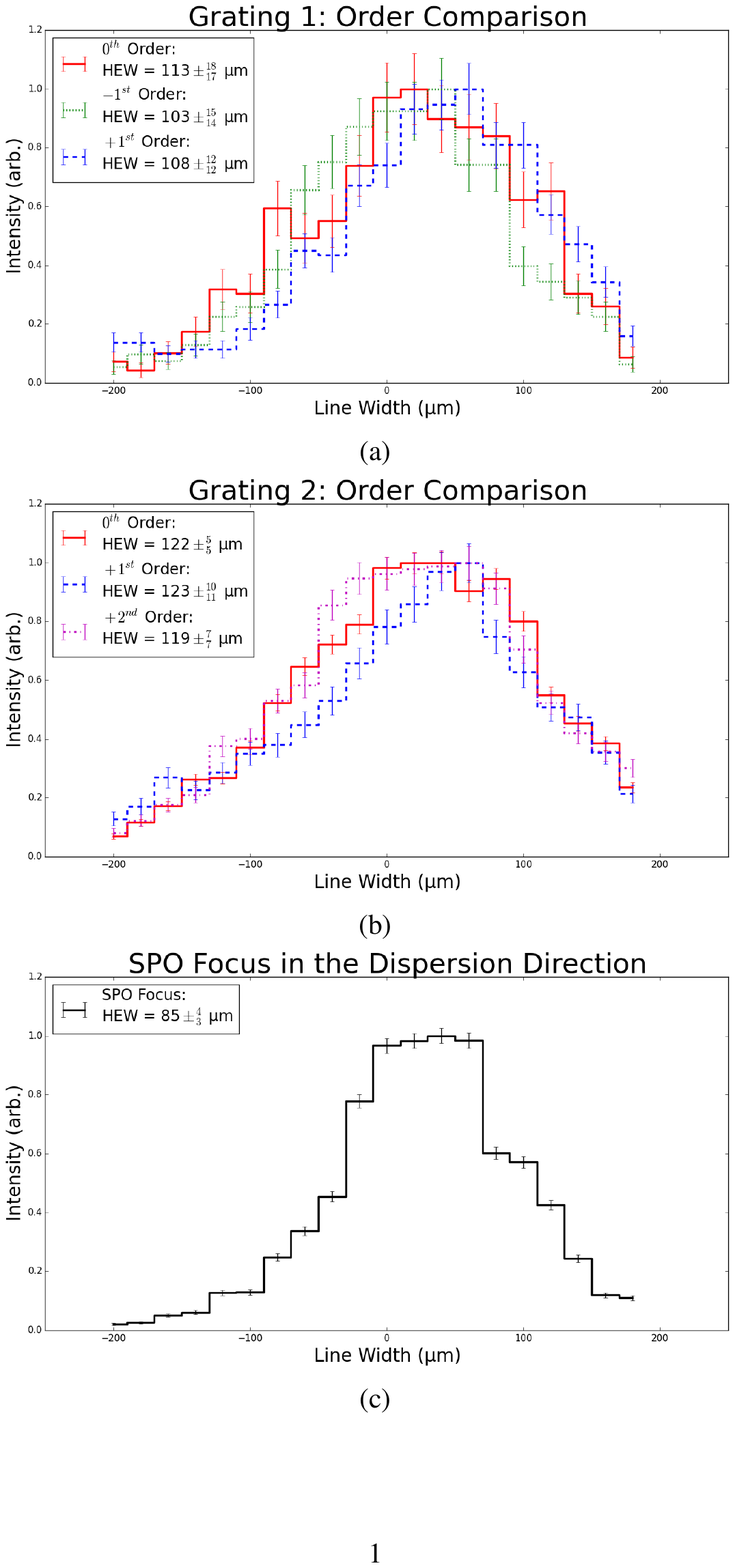}
	\end{tabular}
	\end{center}
	\caption{\label{fig:LSF_comparison} The LSFs of all observed orders and the PSF of the SPO compared side-by-side. (a) The LSFs of imaged Mg K$\alpha$ orders from Grating 1.  Note that because the dispersion direction is reversed from positive to negative orders, any spectral structure in the +1\textsuperscript{st} order Mg K$\alpha$ line should likewise be reversed in the -1\textsuperscript{st} order line. (b) The LSFs of imaged Mg K$\alpha$ orders from Grating 2. (c) The PSF of the SPO focus collapsed in the cross-dispersion direction. Each profile is centered about zero and is normalized to a relative intensity of 1.0.} 
\end{figure}

The LSFs of all measured orders, as well as the \emph{x} extent of the SPO focus with the 63\% mask in place, are shown in Figure \ref{fig:LSF_comparison}. The diffracted order LSFs of Grating 1 and Grating 2 are insets Figure \ref{fig:LSF_comparison}(a) and Figure \ref{fig:LSF_comparison}(b) respectively, while the SPO focus is shown individually in Figure \ref{fig:LSF_comparison}(c). A summary of all measured HEWs, the corresponding 1$\upsigma$ bounds, and the resolutions achieved is presented in Table \ref{tab:HEW_measurements}. As a gentle reminder to the reader, Grating 1 and Grating 2 are blazed in opposing directions, have different facet angles, and are tested at different graze angles (see Sec. \ref{subsec:measurement_configurations} for a detailed summary). Hence, the intensities and LSFs of each measured order should not be expected to be identical from grating to grating. For example, while the Mg K$\alpha$ -1\textsuperscript{st} order was accessible with TRoPIC for both gratings, no concentration of flux was measured at the position of the order for Grating 2. This result is not surprising, given that in the test configuration for Grating 2, the Mg K$\alpha$ -1\textsuperscript{st} order is almost in evanescence and opposite the blaze direction (see Figure \ref{fig:Order_Location_Grating2}). As such, this order is excluded from Figure \ref{fig:LSF_comparison} and no HEW is reported.

The resolving powers of each grating in the current configuration are estimated by dividing the distance dispersed by the observed HEW and reported in Table \ref{tab:HEW_measurements}. In $\pm$1\textsuperscript{st} order for both gratings, we find resolving powers near 400 -- 450, and we report a maximum resolution of R $=$ 800 $\pm$ 20 for Mg K$\alpha$ +2\textsuperscript{nd} order for Grating 2.

\begin{table}
\begin{center}
	\caption{\label{tab:HEW_measurements} The HEWs of the SPO focus and imaged orders.}
	\vspace{1.0em}
	\begin{tabular}{| c | c | c | c | c |}
		\hline
		\multicolumn{5}{| c |}{Measurements of HEW by Diffraction Order} \\
		\hline
		\multirow{2}{*}{Grating} & \multirow{2}{*}{Order} & \multirow{2}{*}{HEW ($\upmu$m)} & \multirow{2}{*}{1$\upsigma$ Errors ($\upmu$m)} & Est. Resolving Power \\ 
		& & & & ($x$/$\Delta x$) \\ \hline \hline
		\multirow{3}{*}{Grating 1} & 0\textsuperscript{} & 113 & \nicefrac{$+$18 }{ $-$17} & ---  \\ 
		& -1\textsuperscript{st} & 103 & \nicefrac{$+$15 }{ $-$14} & 460  $\pm$ 70\\
		& +1\textsuperscript{st} & 108 & \nicefrac{$+$12 }{ $-$12}& 440 $\pm$ 50 \\ \hline
		\multirow{3}{*}{Grating 2} & 0\textsuperscript{} & 122 & \nicefrac{$+$5 }{ $-$5} & --- \\ 
		& +1\textsuperscript{st} & 123 & \nicefrac{$+$10 }{ $-$11} & 390 $\pm$ 30 \\
		& +2\textsuperscript{nd} & 119 & \nicefrac{$+$7 }{ $-$7}  & 800 $\pm$ 20 \\ \hline
		\multirow{1}{*}{SPO Focus -- 63\% Mask} & --- & 85 & \nicefrac{$+$4 }{ $-$3} & --- \\ \hline
		\hline
	\end{tabular}
\end{center}
\end{table}

\subsection{LSFs of Diffracted Orders}
A comparison of the SPO focus, the 0\textsuperscript{th} order focus, and diffracted orders can be used to assess whether the LSF of diffracted orders is aberrated due to defects in the fabricated gratings. A broadening of the LSF due to aberration would impact the ultimate resolution of a spectrometer and could complicate the alignment of multiple gratings, reducing the signal-to-noise ratio in observed orders. To accurately compare the measurements of the SPO focus and grating orders, however, a rigorous accounting of possible errors contributing to the observed LSFs must be made. 

The base LSF of any diffracted order is the PSF of the focusing element. In the present case, however, the SPO focus as reported in Table \ref{tab:HEW_measurements} is not directly comparable to the LSFs of the diffracted orders as the limited grating size (in comparison to the illuminated SPO area) effectively subapertures the optic. This subaperture would be expected to decrease the width of the PSF.  A standard estimate for the reduction in the PSF width for a scatter-dominated X-ray mirror subapertured to an azimuthal range $\theta$ employs a $\sin{\theta}$ scaling law \cite{Cash_1987}. Applying this scaling law would imply a HEW of 51 $\upmu$m for the SPO PSF illuminating the gratings. However, this $\sin{\theta}$ scaling law does not account for the possibility of SPO figure error. Empirical measurements of the SPO focus taken with PIXI during preliminary alignment suggest that the figure error of the optic does make a significant contribution to the width of the SPO focus, and hence the $\sin{\theta}$ scaling is not appropriate for the SPO stack used in this test. As mentioned previously in Sec. \ref{subsec:SPO}, a 9\% SPO mask was employed during preliminary alignment of the optic. As employing the 9\% mask illuminates a significantly smaller azimuthal extent of the SPO stack, this preliminary alignment data is useful in quantifying the expected subaperture effect from the gratings. The best PSF produced by the SPO during preliminary alignment with the 63\% mask ($\approx$ 5.5$\degree$ in azimuth) is measured to have a 135 $\upmu$m HEW. Applying the $\sin{\theta}$ scaling to the focus achieved with the 63\% mask would imply a PSF HEW of 19 $\upmu$m for the 9\% mask ($\approx$ 0.8$\degree$ in azimuth), while the best PSF with the 9\% mask during this same phase of alignment is measured to have a width of 114 $\upmu$m. Thus, we posit that figure error of the SPO plates must also be a contributing factor in the grating-subapertured measurements. Note that as the preliminary alignment data does not have the same focus position used in the test campaign, the SPO PSFs reported here are therefore not directly comparable to the grating test data reported in Table \ref{tab:HEW_measurements}. 

Given the discrepancy between the $\sin{\theta}$ prediction and the measured SPO focus of the 9\% mask during preliminary alignment, we instead estimate the width of the SPO PSF subapertured by the grating during the testing phase by linearly relating the SPO widths measured during preliminary alignment to the azimuthal extents covered by each mask and solving for an azimuthal scale factor $S_{az}$. This relation is then used to solve for the expected width of the grating-subapertured SPO PSF from the 63\% mask data as measured during the test campaign. This method is limited, however, in its ability to account for the radial subaperture of the grating. Each grating is illuminated by 1 -- 2 SPO plates, while the mask data represents the integrated contributions of all 13 plates in the SPO stack (albeit over a limited azimuthal range). Thus, this method does not account for plate misalignments and hence may overestimate the contribution of the SPO to the observed grating LSFs. 

Put mathematically, the scaling method described here calculates the width of the grating-subapetured SPO PSF in the following manner:
\begin{gather*}
\text{PSF Width}_{\text{63\% Mask, Prelim.}} - \text{PSF Width}_{\text{9\% Mask, Prelim.}} = S_{az}(\theta_{\text{63\% Mask}} - \theta_{\text{9\% Mask}}), \\  
\text{PSF Width}_{\text{grating}} = \text{PSF Width}_{\text{63\% Mask, Test}} - S_{az}(\theta_{\text{63\% Mask}} - \theta_{\text{grating}}).
\label{eq:SPO_subaperture}
\end{gather*}

\noindent This method yields an estimate of 76 $\upmu$m for the width of the SPO PSF subaperatured by both gratings, which we take to be the base width of the LSF for the grating orders. 

Next, we compare the 0\textsuperscript{th} order LSF to this estimate of the SPO focus to determine whether there is any broadening of the LSF due to grating figure or a mismatch of the sampled focal plane to the geometrically defined grating focal plane. The measured 0\textsuperscript{th} order focus of both Grating 1 and Grating 2 is inconsistent with the estimated width of the subaperatured SPO focus (76 $\upmu$m). A post-analysis review of the stage positions used during grating measurements shows the diffracted orders of Grating 2 were sampled 2~mm intrafocal of the geometrically-defined grating focal plane, while the Grating 1 data was taken at a distance 6~mm intrafocal of its ideal focal plane. However, a geometric raytrace of the SPO stack demonstrates that for both gratings, the difference in the sampled focal plane and the ideal grating focal plane would not be expected to significantly contribute to the observed width of 0\textsuperscript{th} order. The geometric raytrace propagates 10$^8$ individual rays through a single SPO reflector and grating. Raytrace measurements of the LSF HEW are repeatable at the 1 $\upmu$m level. Thus, the minimum error distinguishable via the raytrace given the estimated 76 $\upmu$m HEW of the subapertured SPO is 12 $\upmu$m r.m.s. Based on the raytrace, the expected r.m.s. contribution to the HEW of the 0\textsuperscript{th} order spot due to the mismatch of the focal plane is 17 $\upmu$m for Grating 1 and is $<$ 12 $\upmu$m for Grating 2. This is consistent with axial (\emph{z}) scans of the SPO focus taken prior to measurements of the gratings, which yield a focus curve predicting an r.m.s contribution of 17 $\upmu$m for a 6~mm intrafocal sampling and a contribution of 7 $\upmu$m for a 2~mm intrafocal sampling of the SPO focus. 

We therefore attribute the observed broadening of 0\textsuperscript{th} order to grating-induced figure error. Lacking figure measurements of the gratings within their mounting structure, we instead employ the same geometric raytrace employed in the previous focal plane study but deform the grating surface from flat into a symmetric ellipsoid characterized by a single radius of curvature $R_{curv}$. This characterization is akin to characterizing the total figure error by a single Zernike polynomial $\mathbb{Z}^{0}_{2}$ 
(defocus) as would be projected onto the format of the grating. The raytraced 0\textsuperscript{th} order spot must also be rotated by the measured angle between the narrow dimension of the grating focus and the horizontal detector axis in order to give the width of the aberrated spot in the dispersion direction rather than raytrace system coordinates.

Reproducing the growth of the estimated grating-subapertured HEW (76 $\upmu$m) to the 0\textsuperscript{th} order spot size reported in Table \ref{tab:HEW_measurements} requires Grating 1 to have a radius of curvature of 4.60 $\times$ 10$^{4}$ mm, corresponding to a peak-to-valley (P-V) measurement of 2.78 $\upmu$m over the 32 x 25 mm grating format. This is outside the P-V tolerance specified for the fused silica wafers used for the fabrication of Grating 1, suggesting that the observed figure error may be attributable to stress induced by the grating mount; however, without an interferometric measurement of a fabricated grating on fused silica, figure error native to the grating itself cannot be ruled out. Grating 2 is found to have a smaller radius of curvature, $R_{curv}$ = 2.75$\times$ 10$^{4}$ mm (P-V: 4.65 $\upmu$m over the grating area). The figure error observed in Grating 2 is not surprising, as the cited flatness specification for the \textless311\textgreater{} wafer used in its fabrication is $<$ 40 $\upmu$m `warp' over its 76.2~mm diameter extent, where `warp' is defined as the sum of the maximum deviations of the wafer above and below the best fit plane.

\begin{figure}[!b]
	\begin{center}
	\begin{tabular}{c}
	\includegraphics[height = 7.5in, clip = true, trim = 2.25in 2.1in 2.25in 1.25in]{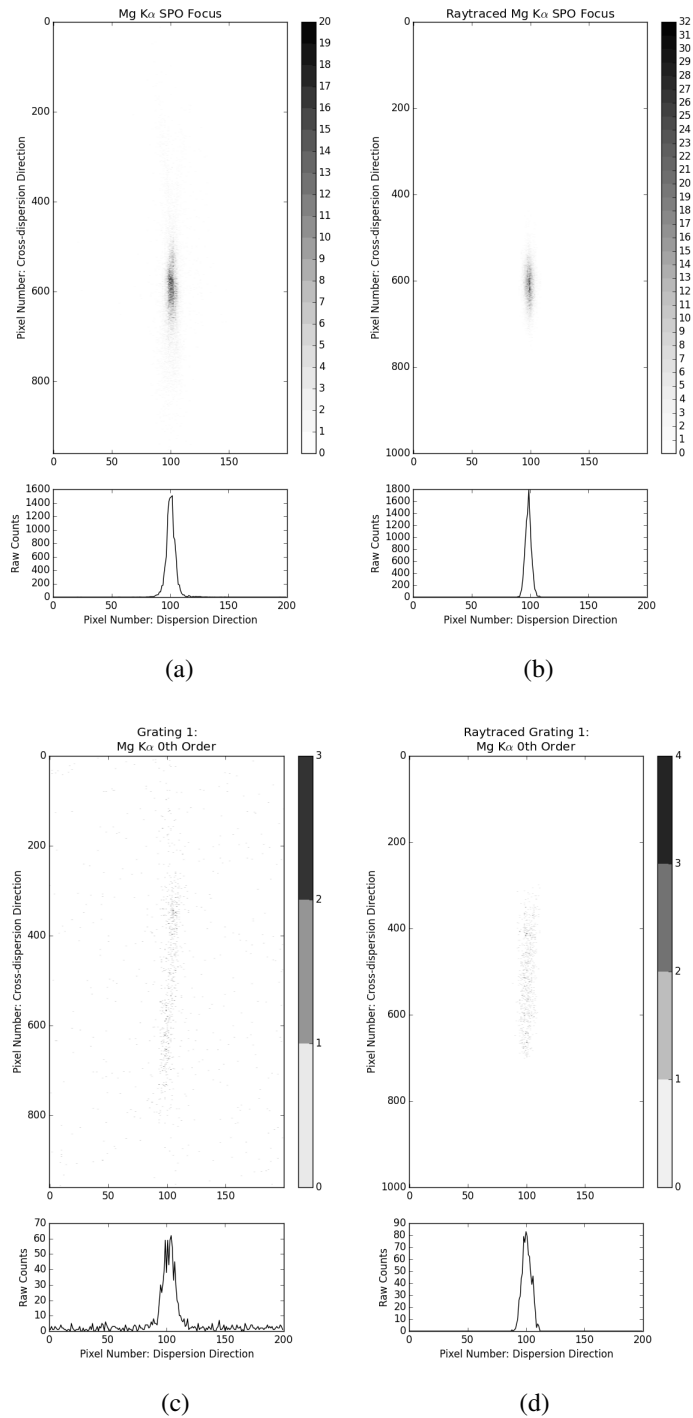}
	\end{tabular}
	\end{center}
	\vspace{-1.0em}
	\caption{\label{fig:raytrace_compare} Measured and raytrace-simulated CCD images of the SPO focus and 0\textsuperscript{th} order focus of Grating 1. Each pixel is 20 $\upmu$m and the images have been summed in the vertical detector dimension to give a line profile. (a) The measured SPO focus with the 63\% aperture mask. (b) The raytraced SPO focus with a simulated 63\% aperture mask. (c) The measured Grating 1 0\textsuperscript{th} order line. (d) The raytraced Grating 1 0\textsuperscript{th} order line with ellipsoidal figure error ($R_{curv}$ = 4.60 $\times$ 10$^{4}$ mm). }
\end{figure}

In addition to quantitatively reproducing the dispersion direction HEWs of the 0\textsuperscript{th} order LSFs for both gratings, the results yielded by the raytrace are also qualitatively similar to the images of the LSFs when binned to a pixel scale identical to the TRoPIC detector format. Figure \ref{fig:raytrace_compare} shows side-by-side comparisons of the measured and raytrace-simulated SPO focus and the Grating 1 0\textsuperscript{th} order LSF. As can be seen in Fig. \ref{fig:raytrace_compare}, the addition of a symmetric ellipsoidal grating figure error also helps account for the observed growth of the 0\textsuperscript{th} order spot in the detector vertical dimension, further supporting the idea that grating figure error is predominantly responsible for the growth of the 0\textsuperscript{th} order LSF relative to the estimated PSF of the SPO stack. 

We also examine via raytrace whether any growth in the LSFs of diffracted orders is anticipated due to a mismatch of the radial convergence of the gratings. As alluded to in Section \ref{subsec:fab_procedure}, the radial pattern of both fabricated gratings is set by the grating pre-master. This pre-master is ruled to match a Wolter-I type telescope with an 8400 mm focal length positioned 250 mm behind the intersection point of the paraboloid and hyperboloid. We model both an idealized radial ruling, in which the ruling is perfectly matched to the geometry of the raytrace system, and the ruling of the pre-master in our raytrace system and find no discernible change in the LSFs of any of the diffracted orders measured in this paper. Thus, we put an upper bound on the contribution of the mismatch of radial ruling equal to 12 $\upmu$m, the minimum r.m.s. contribution to the width of the LSF discernible via raytrace.

Finally, we examine the expected contribution of the finite linewidth of the Mg K$\alpha$ line on the diffracted LSFs. Citrin et al. 1974\textsuperscript{ \cite{Citrin_et_al_1974}} reports an emprically measured full width at half maximum (FWHM) value of 1.10 eV for the Mg K$\alpha$ X-ray emission line fluoresced with an X-ray tube. From this FWHM value, we calculate the expected HEW assuming a Gaussian distributed line and convert this to a width of 4.99 $\times$ 10$^{-3}$ $\angstrom$ centered around the wavelength of Mg K$\alpha$. From the empirically observed dispersion of the gratings of 0.208 $\pm$ 0.001 $\angstrom$/mm in $\pm$1\textsuperscript{st} order, this would yield an expected r.m.s broadening of the line by 24 $\upmu$m in $\pm$1\textsuperscript{st} order and 48 $\upmu$m in 2\textsuperscript{nd} order. 

Table \ref{tab:LSF_errors} gives a full summary of the errors characterized in this section and calculates the total expected HEW of each order by adding the expected r.m.s. error contributions in quadrature with the anticipated width of the grating-subapertured SPO PSF. In our analysis of the diffracted orders, we have assumed that the Mg K$\alpha$ line is adequately described as a single Gaussian despite being composed of two closely-spaced contributions \cite{Klauber_1993}, and that effect of the broadening terms can be accurately described as independent errors summed in quadrature. Given these assumptions, we find total expected HEWs are consistent with the measured HEWs for $\pm$1\textsuperscript{st} orders of both gratings, as can be seen by comparing the measured HEWs stated in Table \ref{tab:HEW_measurements} to the expected LSF HEW given in Table \ref{tab:LSF_errors}. However, the total expected width of the HEW for the +2\textsuperscript{nd} Mg K$\alpha$ line of Grating 2 does not agree with the measurement of this LSF.

\begin{table}
\begin{center}
\caption{\label{tab:LSF_errors} The expected contributions of measurement errors to LSFs of the gratings.}
\vspace{1.0em}
\begin{tabular}{| c | c | c | c | c | c | c |}
\hline
\multicolumn{7}{| c |}{Errors Contributing to the LSFs of Diffracted Orders} \\ \hline \hline
\hline
\textbf{Grating} & \multicolumn{3}{| c |}{\textbf{Grating 1}} & \multicolumn{3}{| c |}{\textbf{Grating 2}} \\ \hline
Order & 0\textsuperscript{th} Order & +1\textsuperscript{st} Order & -1\textsuperscript{st} Order & 0\textsuperscript{th} Order & +1\textsuperscript{st} Order & +2\textsuperscript{nd} Order \\ \hline
Subapertured SPO         				& \multicolumn{3}{c |}{\multirow{2}{*}{76 $\upmu$m}} 			& \multicolumn{3}{c |}{\multirow{2}{*}{76 $\upmu$m}} \\
Focus (est.) 						& \multicolumn{3}{c |}{} 									&\multicolumn{3}{c |}{} \\ \hline

Focal Plane         					& \multicolumn{3}{c |}{\multirow{2}{*}{17 $\upmu$m}} 			& \multicolumn{3}{c |}{\multirow{2}{*}{ $<$ 12 $\upmu$m}} \\
Mismatch (r.m.s.) 					& \multicolumn{3}{c |}{} 									&\multicolumn{3}{c |}{} \\ \hline

$R_{curv}$ Reproducing         			& \multicolumn{3}{c |}{\multirow{2}{*}{4.60 $\times$ 10$^{4}$ mm}} 	& \multicolumn{3}{c |}{\multirow{2}{*}{2.75 $\times$ 10$^{4}$ mm}} \\
0\textsuperscript{th} Order Width 	 	& \multicolumn{3}{c |}{} 									&\multicolumn{3}{c |}{} \\ \hline

Radial Ruling & \multirow{2}{*}{N/A} & \multirow{2}{*}{$<$ 12 $\upmu$m} & \multirow{2}{*}{$<$ 12 $\upmu$m} & \multirow{2}{*}{N/A} & \multirow{2}{*}{$<$ 12 $\upmu$m} & \multirow{2}{*}{$<$ 12 $\upmu$m} \tabularnewline
Mismatch (r.m.s.) & & & & & & \\ \hline

Mg K$\alpha$ & \multirow{2}{*}{N/A} & \multirow{2}{*}{24 $\upmu$m} & \multirow{2}{*}{24 $\upmu$m} & \multirow{2}{*}{N/A} & \multirow{2}{*}{24 $\upmu$m} & \multirow{2}{*}{48 $\upmu$m} \tabularnewline
Linewidth (r.m.s.) & & & & & & \\ \hline
 
Expected & \multirow{2}{*}{113 $\upmu$m} & \multirow{2}{*}{116 $\upmu$m} & \multirow{2}{*}{116 $\upmu$m} & \multirow{2}{*}{122 $\upmu$m} & \multirow{2}{*}{124 $\upmu$m} & \multirow{2}{*}{131 $\upmu$m} \tabularnewline
Totals & & & & & & \\ \hline
\hline
\end{tabular}
\end{center}
\end{table}

We posit that this discrepancy in the width of the +2\textsuperscript{nd} order Mg K$\alpha$ line for Grating 2 arises from the poorly understood contribution of the unpatterned substrate surrounding the grating to 0\textsuperscript{th} order. We are confident in our effort to understand the SPO focus as subapertured by the format of the patterned grating, which forms the base of the LSF for diffracted orders. However, the unpatterned area surrounding the grating would be expected to contribute to 0\textsuperscript{th} order in reflection, effectively creating a different subaperture defined by the overlap of the SPO beam and grating substrate. Subtracting the contribution of the unpatterned substrate from our reported measurement of the 0\textsuperscript{th} order LSF would serve to reduce the measured HEW by as much as 10 $\upmu$m. Quantifying the exact contribution of the unpatterned substrate to the measured 0\textsuperscript{th} order LSF is complicated by the effects of grating figure, which must be a significant contributing factor to the measured 0\textsuperscript{th} order HEW but is degenerate with the effects of a larger subaperture. Therefore, the reported HEW measurement for the Mg K$\alpha$ +2\textsuperscript{nd} order LSF of Grating 2 may be consistent with the broadening expected from the finite width of the line if the 0\textsuperscript{th} order LSF arising from the patterned grating alone is narrower than the 0\textsuperscript{th} order HEW measurement reported, but decoupling the relative contributions of the unpatterned substrate and grating figure cannot be done with the current set of measurements. 

\subsection{Relative Diffraction Efficiencies}
The total number of counts in the LSFs also permit a measurement of the relative intensities of comparable orders. As we do not have knowledge of the total flux incident on the grating, absolute diffraction efficiencies can not be derived from this data. However, the data do permit an estimate of the relative diffraction efficiencies by measuring the total number of counts in a line over a given exposure time. We calculate efficiency ratios for the +1\textsuperscript{st} and -1\textsuperscript{st} orders of Grating 1, and the +1\textsuperscript{st} and +2\textsuperscript{nd} orders of Grating 2. We also calculate a lower bound on the efficiency ratio between -1\textsuperscript{st} and +1\textsuperscript{st} orders of Grating 2 by calculating the efficiency ratio if all recorded photon events in the -1\textsuperscript{st} Mg K$\alpha$ CCD frame were counts in that line. These measurements are reported in Table \ref{tab:relative_efficiencies}. An estimate of the diffraction efficiencies relative to the 0\textsuperscript{th} order reflection is not possible with the current data set, due to the unknown contribution of the unpatterned substrate surrounding the grating to the observed flux.

The measured relative efficiencies reported in Table \ref{tab:relative_efficiencies} demonstrate a clear blaze effect from both fabricated gratings. For Grating 1, -1\textsuperscript{st} order Mg K$\alpha$ is over a factor of two brighter than +1\textsuperscript{st} order, despite -1\textsuperscript{st} order not being located at the blaze position. The blaze effect is more apparent in Grating 2. We observe an almost complete suppression of Mg K$\alpha$ -1\textsuperscript{st} order relative to positive orders, and the Mg K$\alpha$ +2\textsuperscript{nd} line, located roughly at the blaze position, is nearly a factor of two brighter than the +1\textsuperscript{st} order line. The concentration of flux at orders higher than $\pm$1\textsuperscript{st} serves to increase the signal-to-noise ratio in a more spectrally resolved line and has significance for future X-ray grating spectroscopy missions. 

\begin{table}[!t]
\begin{center}
	\caption{\label{tab:relative_efficiencies}The relative diffraction efficiencies for the tested gratings.}
	\vspace{1.0em}
	\begin{tabular}{| c | c | c |}
		\hline
		\multicolumn{3}{| c |}{Relative Efficiencies of Diffracted Orders} \\
		\hline
		Grating & Orders Compared & Relative Efficiency \\ \hline \hline
		\multirow{1}{*}{Grating 1} & -1\textsuperscript{st}/+1\textsuperscript{st} & 2.3 $\pm$ 0.1 \\ \hline
		\multirow{2}{*}{Grating 2} & +1\textsuperscript{st}/-1\textsuperscript{st} & $>$ 43 $\pm$ 10 \\ \cline{2-3}
		& +2\textsuperscript{nd}/+1\textsuperscript{st} & 1.8 $\pm$ 0.04 \\ 
		\hline
	\end{tabular}
\end{center}
\end{table}

\section{Conclusions}
\label{section:conclusions}
Via testing with a SPO at the PANTER X-ray test facility, we have assessed the performance capabilities of two prototype blazed off-plane gratings in the Littrow mounting. The highest resolution demonstrated in these measurements, $R =$ 800 $\pm$ 20, is modest compared to the resolutions required of a future X-ray spectrometer. However, the reported resolution does not represent a systematic limit for the fabricated gratings and could be improved by working at higher order, mitigating the figure error of the tested gratings, and/or improving focus quality of the SPO stack. These measurements have also demonstrated a blaze effect from radially ruled off-plane gratings. Via measurements of the relative intensity of diffracted orders, we have shown that the Littrow mounting can be used to effectively suppress orders far from the blaze position and yield greater throughput in higher orders. The demonstration of this blaze capability has implications for the design of future off-plane X-ray grating spectrometers, as it would serve to concentrate flux on one side of 0\textsuperscript{th} order, reducing the extent of the detector array required to attain the same signal-to-noise ratio and enabling measurements at high order with greater throughput. However, the technique used in the current work to measure grating efficiencies provides no means of assessing the 0\textsuperscript{th} order efficiency with certainty and would be constrained to energies able to be fluoresced by an electron impact source. These limitations motivate the importance of obtaining absolute efficiency measurements at a facility better suited for taking such data, such as a soft X-ray beamline at a synchrotron.

Future measurements of off-plane gratings at beamline facilities like PANTER will greatly benefit from a set of grating aperture masks controlling the illumination of the grating by the focusing optic. The present work employs geometric raytracing in order to estimate the performance of the gratings relative to the SPO PSF. While geometric raytracing is a powerful tool that can be used to better understand expected changes in the performance of X-ray optics, it is only beneficial while employed in conjunction with thorough empirical measurements and can still result in ambiguities. Carefully matching the illuminated portion of the SPO to the illuminated portion of the grating via use of an aperture mask will greatly reduce the need to employ numerical methods to disentangle the contributions of alignment, illumination, and fabrication errors and enable easily comparable measurements of the telescope focus and grating orders.

The inference of a significant figure error contribution to the LSFs of both gratings also motivates the need for direct measurements of the grating surface figure. Gross figure error of individual gratings would substantially reduce the performance of a spectrometer employing many co-aligned gratings, and as such, will need to be minimized in future fabrication efforts. Interferometric measurements of the bare fused silica substrate, after UV-NIL replication, and after the deposition of the X-ray reflective layer will help to elucidate the source(s) of the grating deformation inferred in this beamline test. In the event that the fabricated gratings are found to have a figure error smaller than that needed to explain the observed 0\textsuperscript{th} order deformation, a new grating module which minimizes mounting stress will be devised for future tests.     

In terms of grating fabrication, the next undertaking following this work will be to fabricate large format off-plane gratings via the described fabrication procedure. Fabricating large format gratings will require a new set of nanoimprint molds to be produced. Molds of this size can be made via a commercial process performed by an external vendor. After obtaining these new molds, subsequent gratings can be manufactured over large ($\sim$100 cm$^2$) formats without changing any of the tooling used to complete the fabrication process described in Sec. \ref{subsec:fab_procedure}. The blaze angle afforded by \textless311\textgreater{} Si wafers (29.5$\degree$) is most similar to the grating facet angles baselined for future missions, and a grating with this blaze angle has already been fabricated (Grating 2). Once a large format, blazed off-plane grating has been successfully fabricated, characterizing the figure via interferometric measurements and groove facet profile via atomic force microscopy will provide the information necessary to model the anticipated grating performance. Finally, direct performance testing assessing the throughput and resolving power of such a grating will be a substantial step forward for future spectrometers like \emph{Arcus} and OGRE. 

\acknowledgments 
We would like to thank the anonymous referees for their helpful suggestions and comments. This work was supported by the NASA Roman Technology Fellowship (NNX12AI16G), a NASA Strategic Astrophysical Technology grant (NNX12AF23G), and a NASA Astrophysics Research and Analysis (APRA) grant (NNX13AD03G). We would also like to recognize the support of the University of Iowa Office of the Vice President for Research as well as the College of Liberal Arts and Sciences. The present work benefits tremendously from the University of Iowa Microfabrication Facility and the University of Iowa Central Microscopy Research Facility, a core resource supported by the Vice President for Research \& Economic Development, the Holden Comprehensive Cancer Center and the Carver College of Medicine. Casey DeRoo acknowledges internal funding from the University of Iowa College of Liberal Arts and Sciences. Hearty thanks are due to Wolfgang Burkert and Bernd Budau at the PANTER X-ray Test Facility for their invaluable support during the grating test campaign. 


\bibliographystyle{spiebib}   


\vspace{2ex}\noindent{\bf Casey DeRoo} is a graduate student at the University of Iowa. He received a Bachelor of Arts degree in Physics \& Classical Studies from Concordia College in 2011. His current research interests include optical design, manufacture of optics via microfabrication techniques, and X-ray spectroscopy. He is a member of SPIE.

\vspace{1ex}
\noindent Biographies and photographs of the other authors are not available.

\listoffigures
\listoftables

\end{document}